\documentstyle[12pt,titolo]{article}
\def\k{{k_F}}
\begin{document}
\TITLE{The Nuclear Response \cr in the Isoscalar Channel}
\AUTHOR{R.Cenni, F. Conte and P.Saracco}
\FROM{}
\ABSTRACT{The nuclear response is evaluated in the frame of the 
bosonic 
loop expansion in a purely nucleonic dynamical scheme, which seems to 
be 
reliable in handling those channels where a direct excitation of
a $\Delta$-resonance is not allowed. It is shown that the response 
strongly 
depends upon the effective interaction in the spin-transverse isovector 
channel. New experiments at CEBAF on parity-violating electron 
scattering could further
constrain the form of the effective 
interaction.}
\maketitle

\baselineskip=2.\baselineskip
In previous papers \cite{AlCeMoSa-90,CeSa-94} 
the nuclear charge-longitudinal response function was examined in 
detail
within the theoretical frame of the bosonic loop expansion,
developed in 
refs. \cite{AlCeMoSa-87,AlCeMoSa-88,CeSa-89,CeCoCoSa-92}. 
The same scheme was also successfully applied to the study 
of the energy and momentum dependence of the effective interaction in 
the 
$S=1, T=1$ channels \cite{CeSa-90}.

\begin{figure}[d]
\vskip10cm
\special{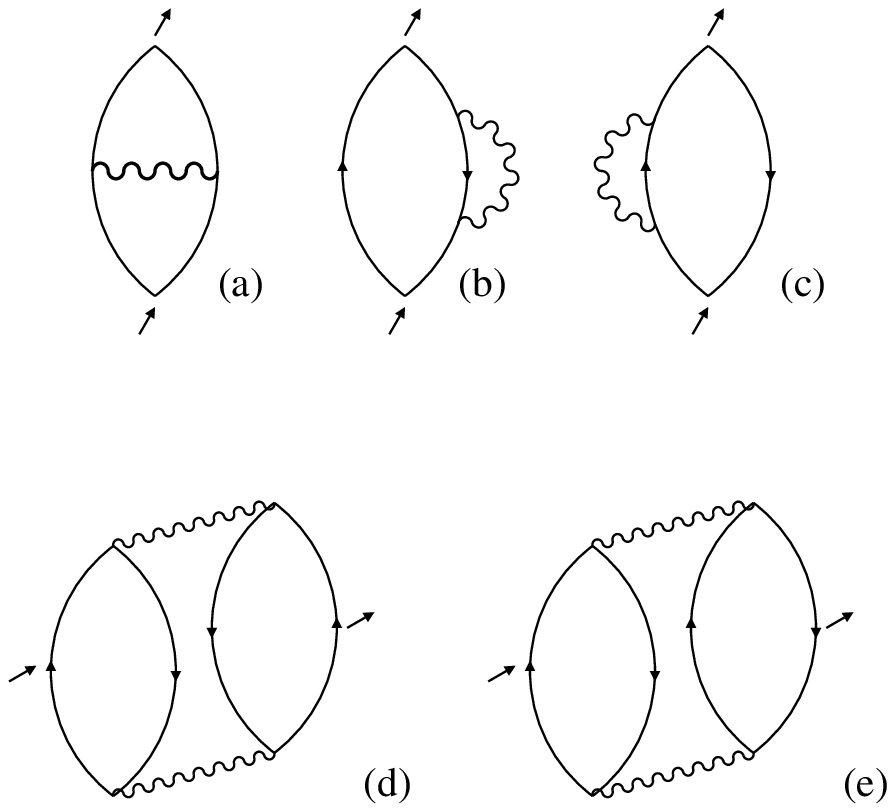}
\caption{\protect\label{fig1}The Feynman diagrams considered in 
present 
work: a): exchange, b) and c): self-energy, d) and e): correlation}
\end{figure}

The key issue of this approach is that a well-behaved expansion can be 
constructed by gathering together all those diagrams which have
the same number of RPA-dressed bosonic loops. 
To properly determine the order of any given Feynman diagram one
should shrink to a point
each fermionic line in the graph, after having RPA-dressed all its
interaction lines: the number of loops so obtained fixes its order.
The bosonic loop expansion for the nuclear response amounts,
at the 1-loop level, to evaluate the diagrams of fig. \ref{fig1}.
We will specify later the interaction we are going to use, 
but, from the beginning, we limit ourselves in the 
present work to nucleonic degrees of 
freedom only. Thus {\em the $\Delta$'s in the ground
state are neglected and the probe cannot allow any $N$-$\Delta$ 
transition}.
The channels expected to be reasonably well described
in this context are then $T=0\ S=0,\ T=0\ S=1$ and $T=1\ S=0$.
We expect instead that the $N$-$\Delta$ transition induced by the 
external 
probe significantly alters the $T=1\ S=1$ channel, as some 
preliminary results already show.

The dynamics considered in \cite{CeSa-94} was the exchange of $\pi,
\rho$ and the transverse component of the $\omega$ plus a residual 
interaction in the two first cases.

The longitudinal propagation of the $\omega$, as well as 
the $\sigma$-meson exchange, are instead neglected. This deserves a 
few 
comments because we know that both provide large contributions in the 
frame of
a mesonic theory.
In different dynamical schemes the effects associated with them can be 
alternatively viewed as coming from other mechanisms, like for instance 
-
in a quark model - the quark exchange effect implied by the
antisymmetrization of a 6--quark bag.
However the overall effect inside a nuclear medium is roughly negligible
independently of the specific dynamical mechanism: this can be 
motivated,
for instance, by the phenomenological analyses of Speth et al.
\cite{SpWeWi-77}, leading
to the conclusion that the Landau parameter in the scalar-isoscalar 
channel -- 
where both mesons propagate -- is compatible with 0, 
thus implying a large cancellation between them.
Alternatively, within a quark model, the inclusion of the 6--quark
configurations requires, to avoid double--counting, to neglect the
longitudinal $\omega$ propagation as well\cite{BuYaFa-89,FaSt-90}.
In a mesonic frame,
where the $\sigma$ is explained in terms of box diagrams\cite{Ho-
81,MaHoEl-87}
(exchange of two isovector mesons with simultaneous excitation of one 
or 
two nucleons to $\Delta$) the cancellation translates into an analogous
one between the longitudinal $\omega$ exchange and the box diagrams. 
Since we are limiting ourselves for the moment to a purely nucleonic 
dynamics, it seemed to us coherent to neglect, together with the 
$\Delta$'s,
those mesons which are in this way associated to them. 
Of course a less crude parametrization of the residual interaction 
in the scalar-isoscalar channel must be pursued, but it cannot be 
directly linked to an existing meson (the 
$\omega$) until we neglect the presence of $\Delta$'s in nuclear 
matter.
Moreover it should be noted that, in the present dynamical model, some
part of the missing short range repulsion is effectively described both
by the values that the ``Landau parameters'' $g^\prime_{L,T}$ actually 
assume
and by the momentum dependence we phenomenologically ascribe to 
the effective
interaction through the cutoffs $q_{c\,L,T}$.

Still a little inconsistency survives in our approach: we included 
indeed the possibility of $\Delta$-h propagation in the RPA-dressing of 
$\pi$ and $\rho$. This is the price we pay to remain in closer contact 
with the phenomenology of the pion propagation in the nuclear medium, 
which must not allow a pion condensation pole. On the other hand, as 
shown in ref \cite{CeSa-94} (and as will be discussed in a subsequent 
work), the $f^\prime$ sum rule is well satisfied in our approach: this 
entails that the quoted inconsistency is truly small.

Coming to the results of ref. \cite{CeSa-94}, the comparison with 
experimental data is in our opinion good, apart from an overall shift of 
the 
data, whose origin was there described in detail. 

However, some problems still survive to fix the input parameters of 
the model. An explicit calculation shows that the depletion of the peak is 
mainly ruled by the correlated $\rho$-meson exchange from diags. (a-c)
of fig. 1.

This is not surprising: in fact, the major contribution comes from the 
short range correlations in that channel,
summarized by $g_T^\prime$ -- the true $\rho$-meson exchange being 
suppressed 
by its high mass -- and in the Landau limit 
\begin{figure}[d]
\vskip10cm
\special{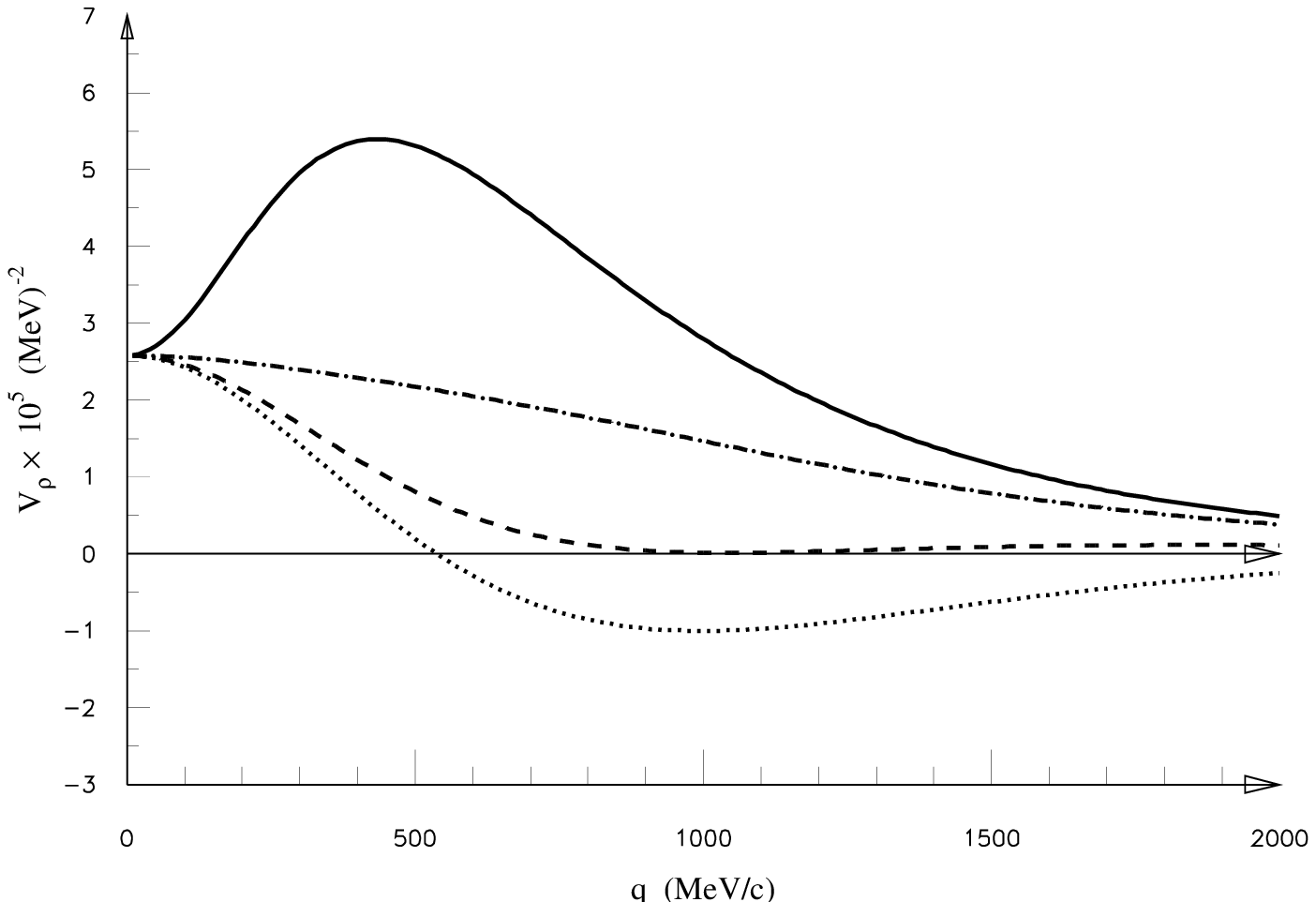}
\vskip1cm
\caption{\protect\label{fig2}The potential $V_\rho$ for different values 
of $q_{c_T}$: from the top to the bottom $q_{c_T}$ = 500 MeV/c, 1000 
MeV/c, 1500 MeV/c, 2000 MeV/c. }
\end{figure}
we know that both $g_T^\prime$ and 
$g_L^\prime$ (the corresponding effective interaction in the pion 
channel)
must coincide. Thus the spatial parts of our diagrams 
receive more or less the same contribution from the $\pi$ and $\rho$
channels, the isospin traces are the same, {\em 
but the traces over the spin matrices in the $\rho$-channel 
are multiplied by a factor 2 with respect
to the pionic one}.

\begin{figure}[d]
\vskip17cm
\hskip-4cm
\special{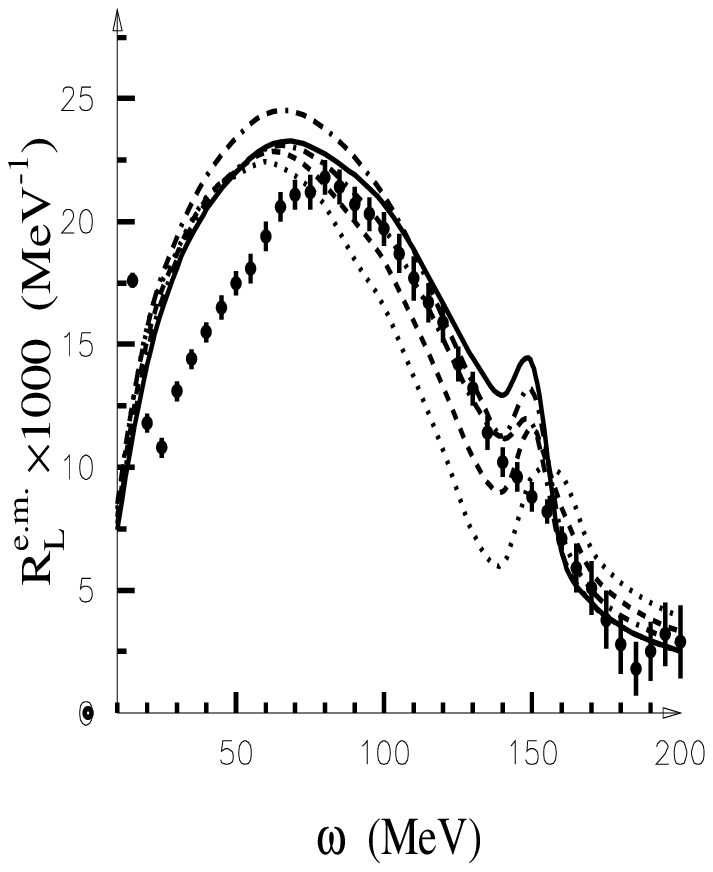}
\hskip7cm
\special{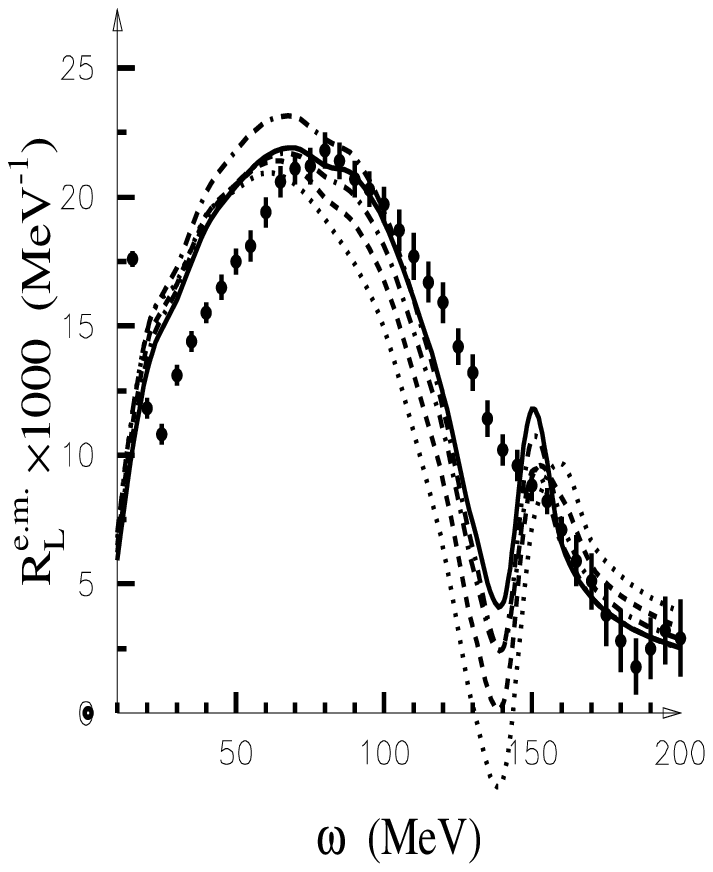}
\vskip-9cm
\caption{\protect\label{fig3}The charge-longitudinal response evaluated 
at different $q_{c_T}$: 1100 MeV/c (dotted line), 1200 MeV/c (dashed 
line), 1300 MeV/c (dash-dotted) and  1400 MeV/c (solid line). 
Calculation is performed with $\k$=1.2 fm$^{-1}$ and $q$=350 MeV/c. 
The dash-dotted-dotted line is the free peak, 
within a non relativistic 
and a relativistic kinematics on the left and on the right, 
respectively.}
\end{figure}

The conclusion of the above simple discussion is that the
dominant contribution to the response comes from the less known 
dynamics.
In fact, in ref. \cite{CeSa-94}, we introduced a parameter ($q_{c_T}$) 
describing
the $\rho$-exchange effective potential in the intermediate range (for 
small
momenta it must tend to $g^\prime_0$ and at higher momenta it is 
expected to
vanish). 
To be explicit, the effective interaction will be written in the form
\begin{equation}
V_\rho(q) = {f^2_{\pi NN} \over m^2_\pi} \left\{ g^\prime_T (q)-C_\rho
{q^2 \over q^2 + m^2_\rho}\right\}v_\rho^2(q^2)
\end{equation}
with
\begin{equation}
g_T^\prime(q)
=1+(g_0^\prime-1)\left[\frac{q^2_{c_T}}{q^2_{c_T}+q^2}\right]^2
\end{equation}
$v_\rho$ being chosen in dipole form with a cutoff of 2500 MeV/c. For 
other details of the calculation we refer the reader to \cite{CeSa-94}.

The behaviour of the potential $V_\rho $ is shown in fig. \ref{fig2}
for $q_{c_T}$ ranging from 500 to 2000 MeV/c.  
We see that the term $g_T^\prime$ is repulsive, 
while the true $\rho$ exchange is attractive, the former being dominant 
at low $q$ and the latter at high momenta. For $q\to \infty$ they must 
exactly cancel, but low values of $q_{c_T}$ mean that the repulsive 
part increases quickly and that the interaction remains always 
repulsive, while very high values of $q_{c_T}$ can allow a change of 
sign of the interaction at a finite value of the momentum.

Now, and this is the central point, different values of $q_{c_T}$, ranging 
from $q_{c_T}$ = 1.1 GeV/c to 1.4 GeV/c can nevertheless satisfactorily 
reproduce
the charge-longitudinal response. In fig. \ref{fig3} we compare indeed 
the 
results of our calculations (according 
to \cite{CeSa-94}) for different values of the parameter 
$q_{c_T}$ with the 
experimental data of Meziani et al..
\cite{Me-al-84,Me-al-85} on $^{12}C$. Here the free quasi-elastic peak 
is evaluated both in a non relativistic and in a relativistic scheme.
The latter, in particular, seems to provide a better 
agreement with the data, but is somehow inconsistent with 
the non relativistic formalism used for the 1-loop corrections.
On the other hand the 
difficulties involved in a fully relativistic calculations seem to be, at 
present, prohibitive and moreover they are expected to produce a few 
percent change on the 1-loop corrections, which are by themselves a 20 
\% of 
the total result. The origin of the instability at 
the edges of the response region has been widely explained in ref. 
\cite{CeSa-94}. 

We note that the different curves display sizable differences but they 
are not decisive - also in view of the corrections stemming
from the terms with $\Delta$-resonances both in the ground or 
intermediate states - to definitely
rule out one of the values of the parameters.

Let us anticipate, before going on, that the corrections quoted above 
are in any case small. They will be discussed in a subsequent paper.

\begin{figure}[d]
\vskip8cm
\special{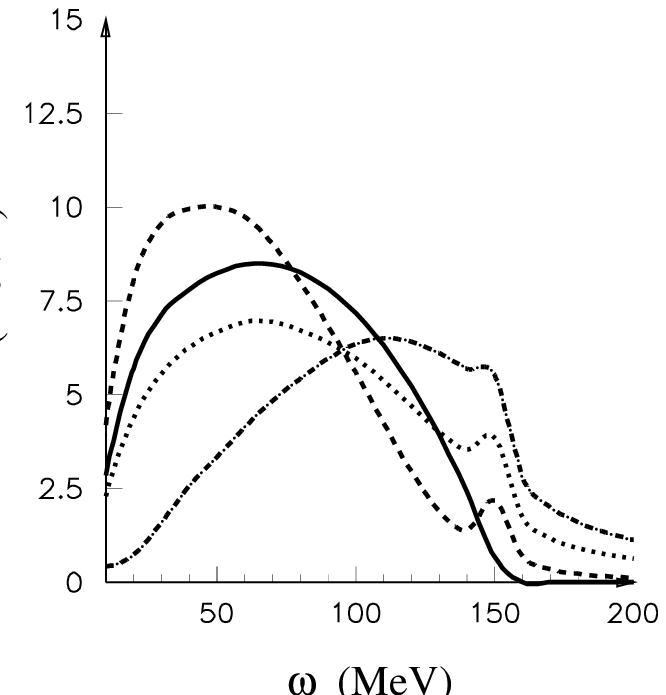}
\hskip7cm
\special{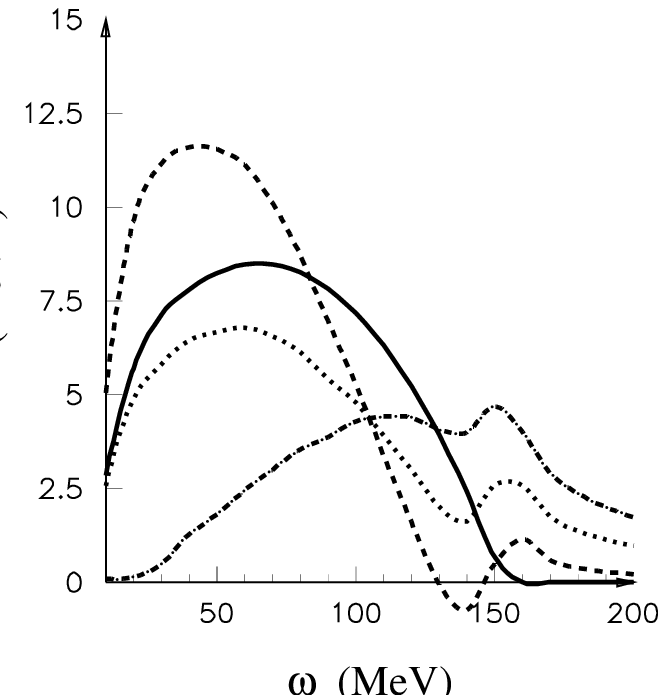}
\caption{\protect\label{fig4}The isoscalar (dashed line) and isovector 
(dash-dotted line) contribution to the charge-longitudinal response for
$q_{c_T}$ = 1400 MeV/c (on the left) and $q_{c_T}$ = 1100 MeV/c (on 
the 
right). The solid line is the non relativistic free Fermi gas and the  
dotted line is 1/2 the sum of the isoscalar and isovector responses.
Calculations are performed with $\k$=1.2 fm$^{-1}$ and $q$=350 
MeV/c.}
\end{figure}

However, when we try to separate the isoscalar and isovector part of the 
charge-longitudinal response the dependence upon
$q_{c_T}$ is emphasized: the separated responses are shown in fig. 
\ref{fig4}
for the 
two extreme cases of $q_{c_T}$ = 1100 MeV/c and  $q_{c_T}$ = 1400 
MeV/c.
The different behaviour between the $S=0\ T=0$ and $S=0\ T=1$ 
channels is
quite evident. The reason resides now in the isospin coefficients.
While in fact the behaviour of the self-energy terms diags. 1b) and 1c)
is {\em exactly} the same in both channel and the correlation terms 1d) 
and 
1e) do not 
differ so much, the coefficient of the exchange
diagram 1a) is 3/2 for the isoscalar part and $-1/2$ for the isovector
one. This diagram was largely responsible\cite{CeSa-94} 
for the cancellation of the 
too high attraction coming from the self-energy (the total result in the 
charge-longitudinal channel being 1). Once the two channels are 
separated the isoscalar one is even more enhanced, while in the 
isovector no compensation arise to the self-energy, the net result being 
a strong depletion.

Clearly the amount of the latter depends upon the details of the model. 
It is however clear, and largely model-independent, that the isovector 
channel is expected to be depleted and that this depletion depends upon 
the strength of the interaction.
Remarkably the smoother isovector response corresponds to a potential 
in the 
$\rho$-channel having some attraction in the intermediate momentum 
range. Noteworthy the potential used by Oset and coworkers [see, e.g. 
\cite{OsSa-87,CaOs-92}] displays exactly this kind of behaviour.

The physical point, however, is that a separation between isoscalar and 
isovector contributions could give us valuable information on the shape 
of 
the effective interaction in the $\rho$-channel. 
The same conclusions hold true also when looking to
the $S=1,\ T=0$ responses, in the spin-longitudinal and transverse 
channels. 

\begin{figure}[d]
\vskip8cm
\special{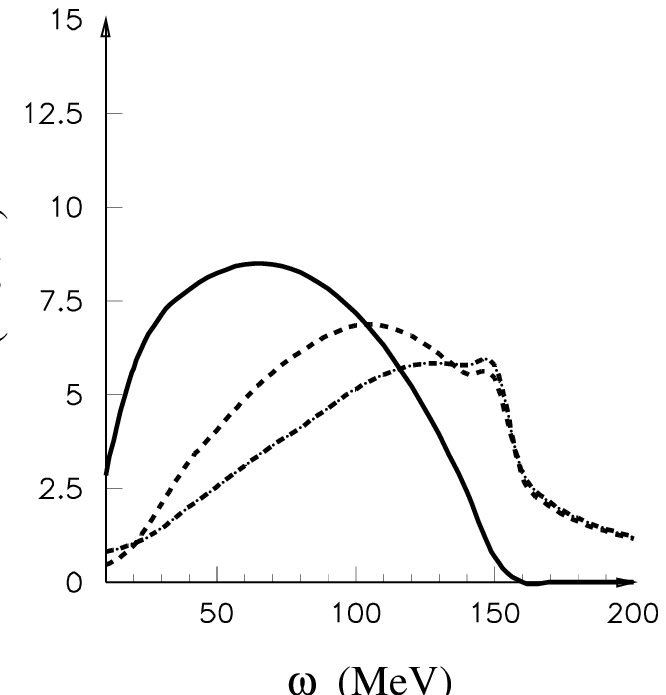}
\hskip7cm
\special{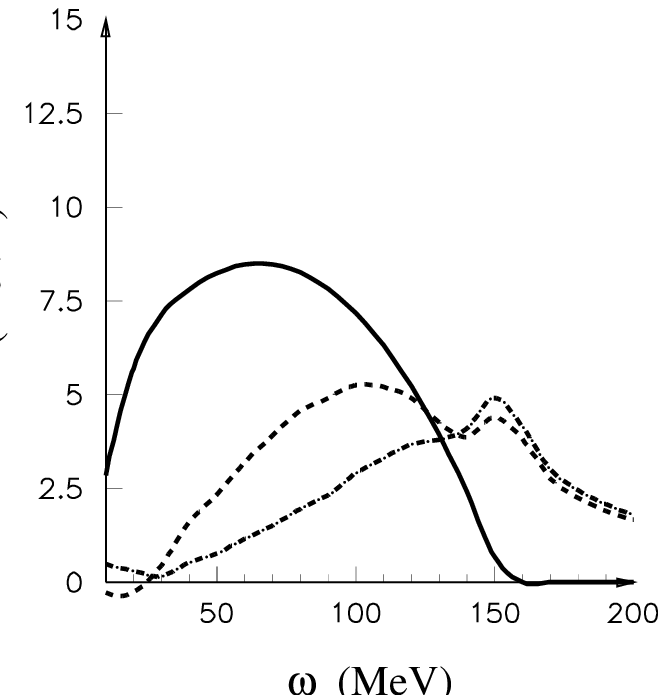}

\caption{\protect\label{fig5}The isoscalar response. Solid line 
represent the Free Fermi Gas, 
the dash-dotted line the isoscalar spin-longitudinal response, the dashed 
line the isoscalar spin-transverse response. Calculations are performed 
with  $q_{c_T}$ = 1400 MeV/c. (left figure) and $q_{c_T}$ = 1100 MeV/c 
(right figure), $\k$=1.2 fm$^{-1}$ and $q$=350 MeV/c.}
\end{figure}
We display in fig. \ref{fig5} the response functions to an isoscalar 
probe, separating the spin-longitudinal and spin-transverse responses. 
The difference between the vector-isoscalar 
channel calculation and the scalar one is again mainly in the spin-isospin 
coefficients; if case also some vertex functions could be different, 
a fact which may only slightly alter the net results.

We again carried out the calculation for 
$q_{c_T}$ = 1100, 1400 MeV/c. Our 
results display in both cases a hardening of the responses,
its amount being again
ruled by the form of the interaction in the $\rho$-channel.
Our outcomes display a sizable difference between isoscalar spin 
longitudinal
and transverse response functions, the former being more depleted in 
the 1p-
1h
response region. 

Notably, the calculations of 
ref. \cite{Fa-94}, performed in the variational frame of the Fermi 
Hyper Netted Chain expansion and
the Correlated Basis Function perturbation theory, are in good 
agreement 
with those of the present work. 

An experience at SATURNE has been recently carried out
\cite{Mo-al-94,Pe-al-94}
where polarized
deuterons are scattered off complex nuclei. By selecting the deuteron 
polarization\cite{Mo-al-92} the experimentalists are able to select 
the $S=1,\ T=0$
channel and, further, to separate the spin-longitudinal and 
spin-transverse
responses. 

Preliminary analyses\cite{Mo-al-94,Pe-al-94} indicate
that the ratio
$R_L/R_T$ for the isoscalar response at $q = 500~{\rm MeV/c}$
 is significantly larger than 1 in the region
inside the quasielastic peak, in
contradiction both with the present work and the one of 
Fabrocini\cite{Fa-94}, and even with older continuum-RPA 
calculations\cite{Sh-al-90}.

These experimental results are however still preliminary and
not yet free from uncertainties on the deuteron 
form factor and on the distortion of its wave function.
It is further conceivable, as stressed by the experimentalists
themselves, that the experimental results could be dominated by
the surface response. 
Only when the analysis will be
completed and one is able to discriminate between surface and volume
effects, the information coming from this experiment
would allow a direct comparison with the present calculation 
and help us in selecting between
the various dynamics which, in our model, provide the same charge-
longitudinal
response, but different effects in the $S=0\ T=0$ and $S=1\ T=0$ 
channels.

On the ground of the present experimental situation the determination 
of 
the volume contribution to the previously quoted responses seems to be 
a 
hard task. We believe however that a larger systematic on a variety of 
nuclei, ranging from carbon to lead, could provide some hints toward the 
desired answer.

Another chance could arise in the future to separate out, instead, the 
two isospin contributions to the $S=0$ channel by means of high 
precision polarized 
electron scattering off nuclei at CEBAF. Could the experimentalists be 
able to perform a Rosenbluth-like separation on the parity violating 
part of the cross section, then the electro-weak longitudinal response 
would become available. This can be expressed, like the e.m. one, as a 
combination of the $T=0$ and $T=1$ contribution but with different 
coefficients. More precisely the parity-violating electro-weak responses 
in the longitudinal channel display an almost complete cancellation 
between isoscalar and isovector contributions at least at the mean field 
level\cite{Do-al-92} , so that only correlations are observed.
Having a model able to provide both the parity-conserving (present 
work)
and the parity violating (work in preparation \cite{CeDoMoSa-95-t}) 
response, and knowing, as it is, the coefficients of the isoscalar and 
isovector contributions, both could be determined.

This would thus offer the opportunity of getting a good knowledge of 
the 
nuclear effective interaction in the otherwise elusive
spin-transverse channel.

\end{document}